\documentclass{aa}
\usepackage{graphicx}

\usepackage{txfonts}

\newcommand{\EWHa}{EW(H$\alpha$)}
\newcommand{\EWHb}{EW(H$\beta$)}

\newcommand{\Ha}{H$\alpha$}
\newcommand{\Hb}{\ifmmode {\rm H}\beta \else H$\beta$\fi}

\newcommand{\hii}{H~{\sc ii}}

\newcommand{\Nii}{[N~{\sc ii}] $\lambda$6584}
\newcommand{\nii}{[N~{\sc ii}]}

\newcommand{\Oii}{[O~{\sc ii}] $\lambda$3727}

\newcommand{\Oiii}{[O~{\sc iii}] $\lambda$5007}

\begin{document}

      \title{What drives the Balmer extinction sequence in spiral galaxies?}

      \subtitle{Clues from the Sloan Digital Sky Survey}

      \author{G. Stasi\'{n}ska\inst{1}
\and
          A. Mateus Jr.\inst{2}
\and
              L. Sodr\'e Jr.\inst{2}
\and
              R. Szczerba\inst{3}
           }

      \offprints{G. Stasi\'nska, \\ \email{grazyna.stasinska@obspm.fr}}

      \institute{LUTH, Observatoire de Meudon, 5 Place Jules Janssen,
                 F-92195 Meudon Cedex, France
\and
                 Departamento de Astronomia, Instituto de Astronomia,
Geof\'{\i}sica e Ci\^encias Atmosf\'ericas da USP, Rua do Mat\~ao 1226,
05508-090,
                 S\~{a}o Paulo, Brazil
     \and
             N.Copernicus Astronomical Center, Rabia{\'n}ska 8,
                 87-100 Toru{\'n}, Poland
     }
           \date{Received ???; accepted ???}
      \titlerunning{What drives the opacity of galaxies?}
\authorrunning{G. Stasi{\'n}ska et al.}


           \abstract{Using  spectra of normal emission line galaxies from
           the First Data Release of the Sloan Digital Sky Survey (SDSS) we
           have investigated the relations between the  extinction
$C({\rm H}\beta)$ as derived from the \Ha/\Hb\ emission line ratio
and various global parameters of the galaxies. Our main findings are
that:
1) $C({\rm H}\beta)$  is linked with the galaxy
spectral type and colour, decreasing from early- to late-type spirals.
2) $C({\rm H}\beta)$ increases with increasing metallicity.
3) $C({\rm H}\beta)$ is larger in galaxies with an older stellar
population.
4) $C({\rm H}\beta)$ is larger for more luminous galaxies.
5) The extinction of the stellar light is correlated with both the
extinction of the nebular light and the intrinsic galaxy colours.
We propose phenomenological interpretations of our empirical
results.
We have also cross-correlated our sample of SDSS galaxies with
the IRAS data base. Due
to the lower redshift limit imposed to our sample and to the detection
limit of IRAS, such a procedure selected only luminous infrared
galaxies. We found that correlations that were shown by other authors
to occur between optical and infrared properties of galaxies
disappear when restricting the sample to luminous infrared galaxies. 
We also found
that the optical properties of the luminous infrared galaxies in our SDSS
  sample are  very similar to those of our entire sample of SDSS
galaxies.
This may be explained by the IRAS luminosity of the galaxies originating
in the regions that formed massive stars less than 1\,Myr ago, while the
opacity of galaxies as derived from the \Ha/\Hb\ emission line ratio is
due to diffuse dust.
  We show some implications of our empirical results on the
determination of global star formation rates and total stellar masses
in normal galaxies.

          \keywords{ Galaxies: spirals -- Galaxies: abundances -- Galaxies:
evolution -- Galaxies: ISM -- Galaxies: stellar content  -- Galaxies:
ISM: dust, extinction  }
}

          \maketitle

\section{Introduction}

Accounting for the presence of dust  is paramount
for our understanding of the constitution and evolution of galaxies.
Indeed, dust  modifies the light we receive from galaxies by both
dimming it and modifying its colour. The determination of the
bolometric luminosity of galaxies, the description of their stellar
populations using galaxy colours, the estimates of the star formation
rates using observed fluxes in either the emission lines or
ultraviolet continua, all depend on a correction for the effects of
dust. Dust in itself is also an important constituent of galaxies, not so
much by its mass but essentially by the effects it has on the thermal
balance of the interstellar medium  and on the
formation of H$_2$ molecules, which has
important consequences on the efficiency of star
formation (e.g. Omukai \cite{O00}, Hirashita \& Ferrara \cite{HF02}).

Yet,
the question of dust opacity of galaxies is still a subject of
strong debate.
Earlier, the presence of dust in galaxies was essentially
inferred from its impact on the observed distribution of stellar light. The
spectacular extinction in the edge-on Sombrero galaxy was
attributed to dust located in its plane.  The
mottled aspect of the arms in face-on spiral galaxies  was attributed to
higher extinction in zones of star formation.
Such observations developed the view that dust is a
common constituent of spiral galaxies.  In lenticular galaxies,
dust lanes have been noticed  (Sandage \cite{S61}).  Studies using
various techniques have led to the conclusion that the average
extinction
decreases following the sequence late-type spirals --
early-type spirals -- lenticulars and ellipticals (see review by Calzetti
\cite{C01}).
Inclination tests measure the
dependence of galaxy disk surface brightness on inclination
(e.g. Valentijn \cite{V94}
and references therein). This method is statistical by nature and
strongly depends on sample selection, as discussed by Valentijn.
Another method is to consider multiwavelength data of galaxies and solve
simultaneously for the intrinsic colours of the stellar
populations and for the reddening. The results of this method strongly
rely on the adopted dust distribution models, as emphasized by Witt
et al.\,(\cite{WTC92}), Bianchi et al.\,(\cite{BFG96}), and Witt \&
Gordon\,(\cite{WG00}). Infrared
data from the IRAS survey provide detection of dust through its emission
in the infrared. However, the interpretation of the infrared dust
emission in terms of dust content is not straightforward (see e.g.
Sauvage \& Thuan \cite{ST94}, Dale \& Helou \cite{DH02}).
A different approach is to study the transparency of
galaxies with respect to background light (see Keel \& White \cite{KW01}
and
references therein for methods using one nearby
background galaxy, and Gonz{\'a}lez et al. \cite{GAD98}
for methods using distant galaxies as
background sources). Unfortunately, this method suffers from
extremely low statistics.

  Another extinction indicator that can be used in galaxies
presenting emission lines is the observed Balmer decrement. As is
known, the intrinsic
intensity ratios of hydrogen recombination lines in ionized nebulae
have a negligible dependence on the emission conditions (Osterbrock
\cite{O89})
so that the observed values are a consequence of the
dust extinction being different at the relevant
wavelengths. This has for example been
used in the influential work by Calzetti et al.\,(\cite{CKS94}) to
determine an
empirical extinction law for starburst galaxies.
However, few studies have used this method to investigate the relation
between  extinction and overall galaxy type (e.g. Hubble type). In
order
to do this, one needs spectra of galaxies spanning
the entire range of Hubble types. The atlas of Kennicutt (\cite{K92})
provided such a data base for a limited sample of galaxies. Among
these galaxies, 15 were considered by  Sodr\'e \&
Stasi\'nska (\cite{SS99}, hereinafter SS99) to be normal emission line
galaxies, and these
authors showed that the extinction at \Hb\, as measured by the \Ha/\Hb\
ratio
   (in the remaining of the paper
  we will call it the \emph{Balmer extinction})
decreases steadily from early-type to late-type spirals.  Later, using
the Nearby Field Galaxy Survey (NFGS) of Jansen et
al. (\cite{JFFC00a}, \cite{JFFC00b}),
which provided adequate data for about 100 galaxies,
Stasi\'nska \& Sodr\'e (\cite{SS01}, hereinafter SS01)
showed that redder galaxies have
larger  Balmer extinction. The Sloan
Digital Sky Survey (SDSS), which aims at obtaining spectra
of $10^{6}$ galaxies in the nearby Universe, provides a wonderful
opportunity to study this question in more detail.

The present paper makes use of the observations from the First Data
Release (DR1; Abazajian et al. 2003, see also Stoughton et al. 2002)
of the SDSS to  study the
relation between the  Balmer extinction and
other global properties of the galaxies. Section 2 explains the
selection of the observational
sample and quantities used in the analysis.
  Section 3 presents  the relation between  Balmer extinction and other
galaxy parameters. Section 4 presents an interpretation of our
results. In Section 5 we cross-correlate the galaxies
of our sample with galaxies detected by IRAS, looking for additional
clues on the origin of the observed extinction.
Section 6 summarizes the main conclusions of this work and outlines 
some prospects.

\begin{figure}
\centerline{\includegraphics[width=8cm]{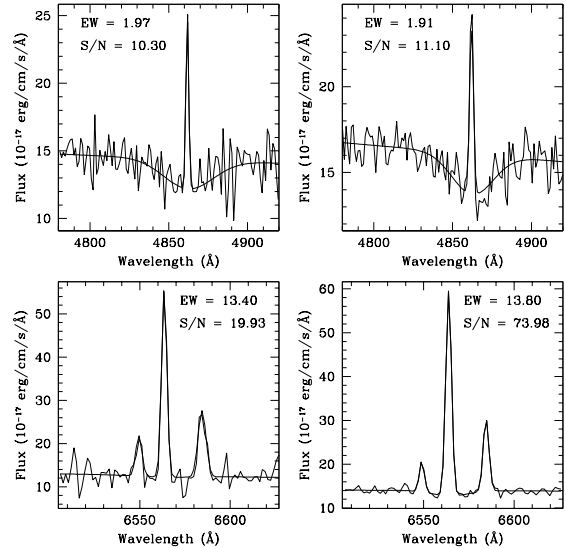}}
\caption{Examples of two fits using our code to measure the equivalent
widths and fluxes of H$\beta$ (top panels) and
H$\alpha$ (bottom panels). The left and right figures are for
spectra with signal-to-noise ratio in $g$-band of 5.9 and 12.4,
respectively. We also show the equivalent widths (EW) in \AA\ and
signal-to-noise ratio (S/N) of the measured emission lines.}
\label{ews1}
\end{figure}

\section{The observational sample}

  \subsection{Measurements of emission line parameters}

The equivalent widths of H$\alpha$
and H$\beta$, \EWHa\ and \EWHb\ respectively,  measured by the SDSS team
and
available in the survey database, do not account for underlying
absorption, which may be significant for objects with faint emission
lines. Consequently, we have developed a code for measuring
simultaneously the emission and absorption features of these Balmer
lines (see Mateus \& Sodr\'e \cite{MS04} for additional details). For
H$\beta$
the code fits
two Gaussian functions, one in emission and the other in absorption,
allowing for different centroids and widths for these features.
For H$\alpha$ we fit, additionally, two other Gaussian functions
to the emission lines of \nii$\lambda$6548 and \nii$\lambda$6584.
The continuum is estimated through a robust linear adjustment considering
two wavelength regions on the sides of the lines. In Fig. \ref{ews1} we
show  examples of the fits obtained for H$\beta$ (top panels) and
H$\alpha$ (bottom panels) for two spectra with a signal-to-noise
ratio in the $g$-band of 5.9 and 12.4, respectively from left to right of
the figure. Note that these are among the worst cases in the sample
selected for this study (see Section 2.2).
This procedure allowed us to determine, for each spectrum,
the fluxes
and equivalent widths resulting from the absorption and emission
features of H$\alpha$ and H$\beta$, as well as their errors and a
signal-to-noise ratio (S/N) for each fit.

\begin{figure}
\centerline{\includegraphics[width=8cm]{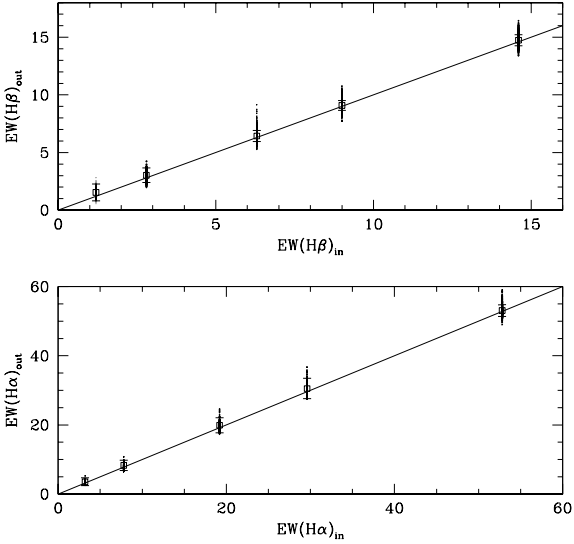}}
\caption{Results of the simulations for a spectral dispersion of 1.0 \AA\
and   typical noise. The input equivalent widths are based on the
results of the spectrophotometric model of
Barbaro \& Poggianti (\cite{BP97}) for different spectral types:
Sa, Sb, Sc, Sd and `Extreme' case with constantly increasing star
formation
rates.
These equivalent widths are compared with the values recovered by our code.
The large squares are the mean values of the output equivalent widths,
with the
error bars showing their dispersion.}
\label{sim1}
\end{figure}

We have made many simulations to verify the reliability of these
fits and the presence of any bias in the results. We simulated
absorption  plus emission lines, with realistic noise, and compared the
input
line
parameters with the outcome of our fitting software. The initial
parameters (essentially the equivalent widths of the components in
emission and absorption of H$\alpha$ and H$\beta$ lines) used in
these simulations were based on the results of the
spectrophotometric model of Barbaro \& Poggianti (\cite{BP97}), for
different spectral types. For each out of 5 types, we simulated 1000
spectral
regions at H$\alpha$ and H$\beta$ lines, varying the amplitude of
the noise, as well as the linear spectral dispersion (0.5, 1.0 and
2.0 \AA). Then, we applied our code to compare the equivalent widths of
the H$\alpha$ and H$\beta$ features. In Fig. \ref{sim1} we show a
typical result obtained from our simulations, where the initial
equivalent widths of H$\alpha$ and H$\beta$ (input parameters) are
compared with their respective values measured by our code.
We concluded that for S/N high enough (larger than $\sim$ 10) the results
recovered by our software are, within the errors, consistent with the
input values and do not depend on the dispersion we used.

In this study we use our own  measurements of the equivalent widths
of the hydrogen emission
lines, and the measurements available in the SDSS database for the
remaining
lines.

\begin{figure}
\centerline{\includegraphics[width=5.13cm]{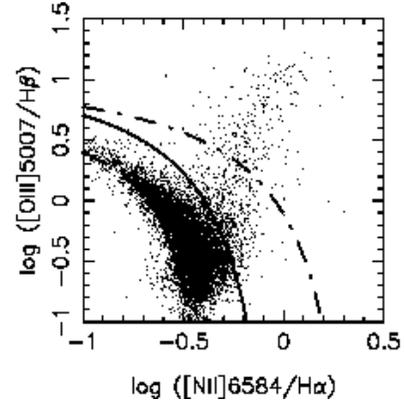}}
\caption{Emission line diagnostic diagram to separate normal
galaxies from galaxies hosting an AGN. The points represent the $10822$
galaxies from the SDSS that have been selected according to the
criteria described in Sec. 2.1. Solid line:  the  curve of Heckman \&
Kauffmann (2003); all the galaxies below that curve
(9840 objects) constitute our sample of normal galaxies.
Dot-dashed line: the theoretical curve of Kewley et al. (2001)
above which galaxies are dominated by an AGN; dashed line: empirical
lower bound of normal galaxies (see  Sect. 2.2).
}
\label{fig1}
\end{figure}

\subsection{The selection procedure}

The spectroscopic data analyzed in this work were extracted from the
SDSS main galaxy sample available on the First Data Release
(Abazajian et al. \cite{A03}).
This flux-limited sample consists of galaxies with $r$-band Petrosian
magnitudes $r \le 17.77$ and $r$-band Petrosian half-light surface
brightnesses $\mu_{50} \le 24.5$~mag~arcsec$^{-2}$,
comprising a total of 113199 galaxies. We first selected all
galaxies with spectra obtained by the survey and with a redshift
confidence $\ge 0.7$.
This selection  resulted in a first sample with 110913 galaxies.
Following the conclusions of Zaritsky, Zabludoff \& Willick
(\cite{ZZW95}),
we have used a redshift limit of $z \ge 0.05$ to minimize the effects
of the aperture bias in our sample, since we are interested in the
global spectral properties of the galaxies. We have also rejected
galaxies with an average spectroscopic signal-to-noise ratio smaller
than $5$ in the photometric $g$ passband. These criteria result in a
selection of 89174 spectra of galaxies.

For these objects, we have applied our code to measure the fluxes and
equivalent widths of the emission Balmer lines, obtaining 40672
spectra with these parameters measured with S/N $>$ 1.
Following the previous discussion on the H$\alpha$ and H$\beta$
measurements, we consider only galaxies with high S/N for their
emission lines, equal to or larger than 10, and
measured \EWHa\ and \EWHb\ larger than 1~\AA. For galaxies with
  strong emission in the Balmer lines
(where the absorption equivalent width is much smaller than the
emission equivalent width), our emission
equivalent widths are very similar to those in the SDSS database.
With these criteria, the sample comprises $11066$ spectra. In the case of
duplicate spectra,  we removed the ones with lower S/N in the $g$ band. The
resulting sample contains 10854 objects.

We then excluded from the sample  the galaxies whose spectra are
contaminated by non-stellar activity. For this, we used the
\Oiii/\Hb\ vs. \Nii/\Ha\ diagram introduced by Baldwin, Phillips \&
Terlevich (\cite{BPT81}) and Veilleux \& Osterbrock (\cite{VO87}).
The resulting diagram for our sample is shown in Fig. \ref{fig1},
where we only plot the objects with measured fluxes for all emission
lines ($10822$ objects).
The dot-dashed line is a limit defined by Kewley et al. (\cite{KWDL01}),
above
which an active galactic nucleus (AGN) dominates the emission-line
spectrum.
The solid curve shows
the empirical limit proposed by Heckman \& Kauffmann (\cite{HK03}),
below which star formation dominates the emission-line spectrum.
Galaxies below this curve will be referred to as normal galaxies.
The dashed line delimits the
narrow strip where normal galaxies are found in this diagram.
Note that this strip has roughly the same location as
the sequence drawn by individual giant \hii\ regions in galaxies
(Mc Call et al. \cite{MRS85}, van Zee et al. \cite{vZ98}).
Thus, in our selection we can distinguish $9840$ normal galaxies
and $982$ AGNs.
In the following, our sample will contain only these $9840$ normal
emission line galaxies, to avoid any interference of active nuclei
in our interpretation of the data.

\begin{figure}
\centerline{\includegraphics[width=8cm]{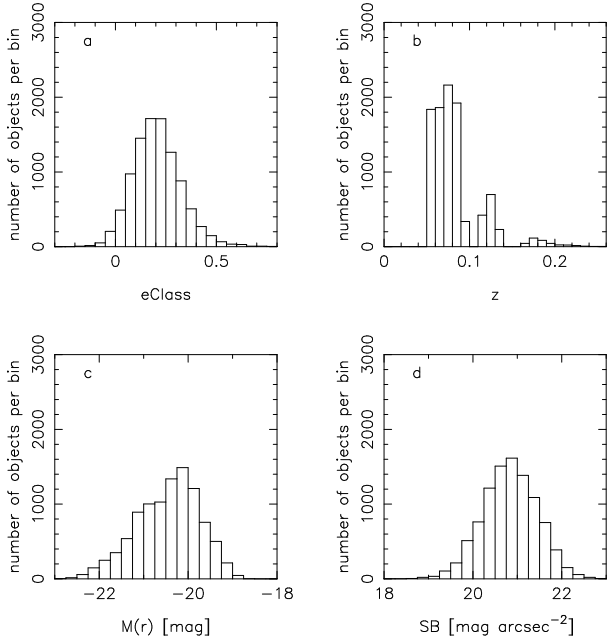}}
\caption{Histograms of general properties of our sample of normal
galaxies.
}
\label{fig2}
\end{figure}

\begin{figure}
\centerline{\includegraphics[width=8cm]{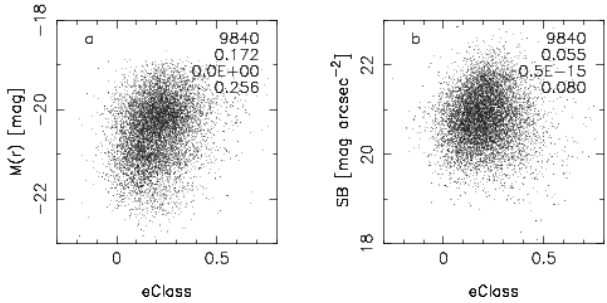}}
\caption{Some global properties of the galaxies of our sample
  as a function of their ``spectral type''. The top number on the right
is the
total number of data
points (including a few individual points that may lie outside the
panel frame). Below are: the value of Kendall's $\tau_{K}$, then the
associated
probability and  finally the Spearman rank correlation coefficient
$r_{S}$ (see Sect. 2.3).
}
\label{fig3}
\end{figure}

\begin{figure*}
\centerline{\includegraphics[width=17cm]{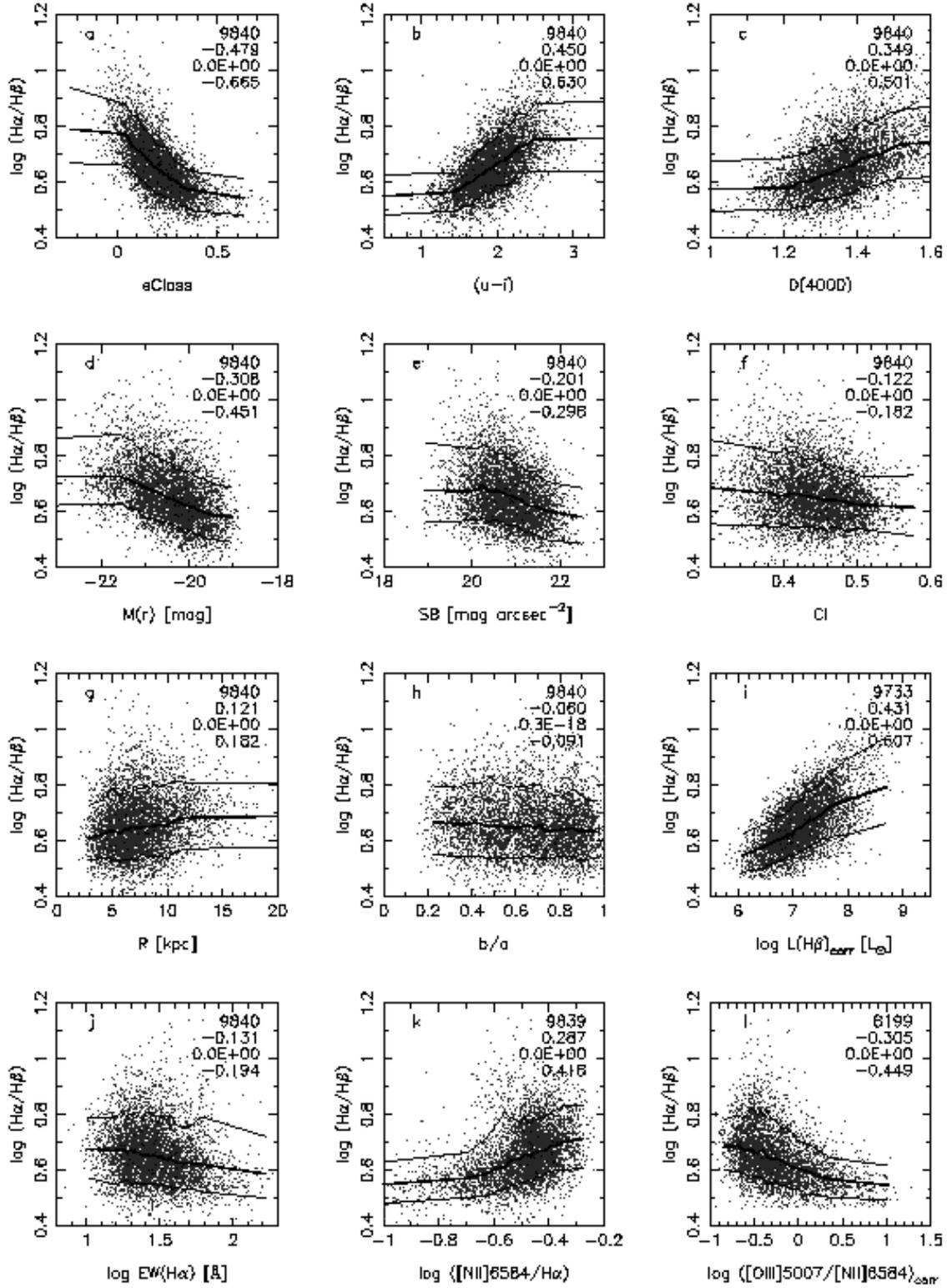}}
\caption{The relation between log (\Ha/\Hb) and various properties of
the galaxies. The meaning of the numbers in the upper right is the
same as in Fig. 5. The thick curve
represents the median value of log (\Ha/\Hb).  The thin curves
delimit the zone containing 80\% of the data points (see Sect.
3.1).
}
\label{fig4}
\end{figure*}

\subsection{Overall properties of our sample of normal galaxies}

With 9840 objects, we have over
80 times more galaxies at our disposal than in the SS01 study based on
the NFGS sample (Jansen et al. \cite{JFFC00a}, \cite{JFFC00b}),
and  over 600 times more
than in the SS99 study based on the Kennicutt (\cite{K92}) atlas.  Note
also that at the
spectral resolution of the SDSS, which is about 3~\AA,  it is much
easier to separate emission from absorption at \Hb\ and \Ha\
than in the NFGS (spectral
resolution  6~\AA) and in the Kennicutt (\cite{K92}) atlas (spectral
resolution 12--25~\AA). Therefore we are able to explore with
confidence galaxies with quite weak \Hb\ emission.

The global characteristics of the selected
sample are shown in  Fig. 4. Panel $a$ shows the histogram of the
galaxy spectral classification parameter \texttt{eClass}  as provided by
the
SDSS for DR1. This spectral classification parameter is based on
a principal component analysis method (Connolly et
al. \cite{CSB95},
Connolly \& Szalay \cite{CS99}). Note that the spectral
classification of galaxies by Connolly et al. (\cite{CSB95})
is slightly different from
the one used in SS99 who removed from the analysis the wavelength regions
containing emission lines.  We will nevertheless refer to the
parameter \texttt{eClass} as to a ``galaxy spectral type''. Note
that  \texttt{eClass} is strongly correlated with the galaxy colour
$(u-i)$. The values of
  \texttt{eClass},
  range from about -0.2
to 0.6 from early- to late-type in our sample of galaxies. Panel $b$
shows the
distribution of redshifts $z$. Some SDSS spectra have residual sky lines
that
may affect our H$\alpha$
and H$\beta$ measurements. Despite the mask array contained in each
spectrum file that can be used to remove any inconvenient features,
like bright sky lines, we have chosen an alternative way by excluding all
objects in which the bright sky lines at $5578.5$ {\AA} and $7246.0$ {\AA}
fall on the H$\alpha$ and H$\beta$ line regions, which explains the
bizarre aspect of the redshift distribution
of this sample. Panel $c$ shows the
distribution of galaxy absolute magnitudes in the SDSS
photometric $r$ band, $M(r)$. This magnitude is utilized because
  K-corrections are
modest, the radiation is produced mainly by the older stars that dominate
the stellar mass, and uncertainties in Galactic reddening make little
difference to the inferred galaxy magnitude (Strauss et al. \cite{SWL02}).
To compute $M(r)$ from the observed $r$
magnitude given by the SDSS, we assumed a standard cosmology with
$\Omega_{M}$ = 0.3, $\Omega_{\lambda}$ = 0.7, and
$H_{0}$ = 75~km~s$^{-1}$~Mpc$^{-1}$ and applied a K-correction
(computed with \texttt{kcorrect v1\_11}; Blanton et al. \cite{B02}).

Because of our restrictions on the quality of the spectra, we miss the
less luminous galaxies of the SDSS (compare e.g. the histogram of the
$M(r)$ values in Blanton et al. \cite{BDE01}).
Finally, panel $d$ shows the distribution of the galaxy surface
brightness, ${SB}$ (in mag arcsec$^{-2}$), defined within the radius
containing
50\% of
the Petrosian flux for each band, \texttt{petroR50}.
Note that for this study, we do not work with a complete
volume- or magnitude- limited sample, so that the distribution of
galaxies among the various classes does not  correspond to
the true distribution in the local universe. Also, note that,
because of the conditions imposed on
the quality of the spectra, the redshifts in our sample do not exceed
a value of 0.25.

\begin{figure}
\centerline{\includegraphics[width=5.cm]{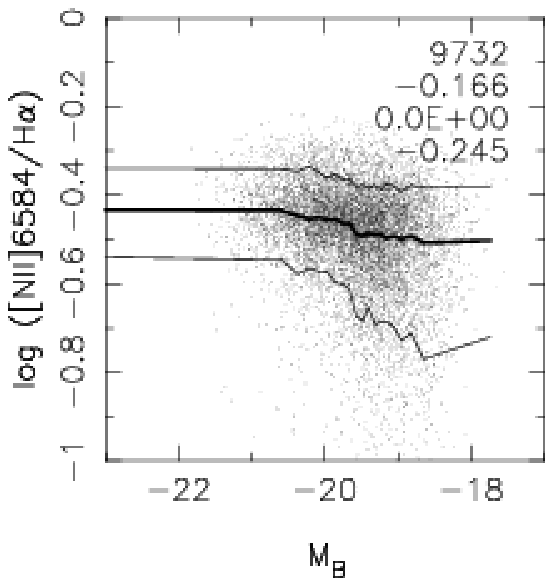}}
\caption{The relation between the metallicity indicator \Nii/\Ha\ and
the absolute magnitude $M_{\rm B}$ in the galaxies of our sample. Same
layout as for the panels in Fig. 6.
}
\label{fig5}
\end{figure}

\begin{figure*}[t]
\centerline{\includegraphics[width=16cm]{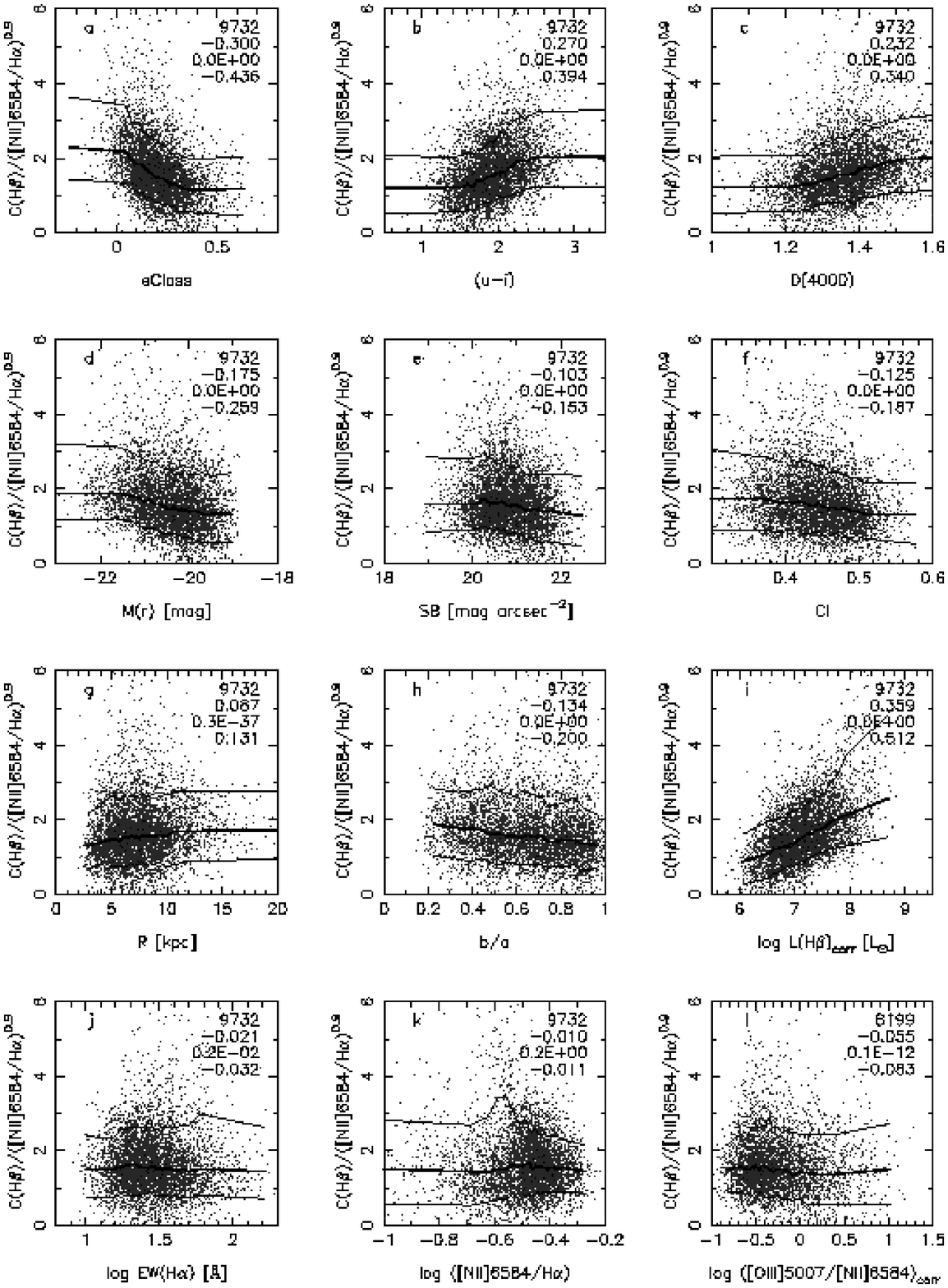}}
\caption{Same diagrams as in Fig. 6 except that the ordinate is
$C({\rm H}\beta)$/(\Nii/\Ha)$^{0.9}$ instead of log (\Ha/\Hb) (see
Sect. 3.3)
}
\label{fig6}
\end{figure*}

Some of the overall characteristics of galaxies are in fact
correlated. For example, galaxies of earlier spectral types tend to
be more luminous, as shown in panel $a$ of Fig. 5 which plots $M(r)$
vs. \texttt{eClass}.
The top number on the right corresponds to the total number of data
points (including a few individual points that may lie outside the
panel frame). Next, are given the results of non parametrical tests:
from top to bottom, the value of Kendall's $\tau_{K}$, the associated
probability (a large value of this probability indicates no correlation)
and
the Spearman rank correlation coefficient $r_{S}$
(see Press et al. \cite{P92}). Thus, we see that indeed $M(r)$ is
correlated with \texttt{eClass}. On the other hand, the galaxy surface
brightness $S\!B$ is virtually independent of \texttt{eClass} (panel $b$).

\section{ Balmer extinction of normal galaxies}

\subsection{Definition of the diagnostic plots}

  The value of the Balmer extinction
  is derived from the observed ratio of
\Ha\ and \Hb\ line intensities
using the relation
\begin{equation}
C({\rm H}\beta) = \frac{1}{f({\rm H}\beta)-f({\rm H}\alpha)}
{\rm log} \frac{I({\rm H}\alpha)/I({\rm H}\beta)}
{I^{0}({\rm H}\alpha)/I^{0}({\rm H}\beta)}
\end{equation}
where  ${I^{0}({\rm H}\alpha)/I^{0}({\rm H}\beta)}$
is the intrinsic intensity ratio of these two lines and
$f({\rm H}\beta)-f({\rm H}\alpha)$ is equal to 0.335.
The value of ${I^{0}({\rm H}\alpha)/I^{0}({\rm H}\beta)}$ is 
essentially insensitive to the physical conditions of the gas, being 
3.03
  at a temperature of 5000~K, 2.86 at 10000~K, and 2.74 at 20000~K 
(Osterbrock \cite{O89}).
Thus, there is a practically linear relation between  $C({\rm
H}\beta)$ and
log\,($I({\rm H}\alpha)/I({\rm H}\beta)$)
(in the following  simply noted log (\Ha/\Hb)).

In Fig. 6, we plot the value of log (\Ha/\Hb) as a function of
various parameters characterizing the galaxies.  In all the panels,
the presentation is the same. The meaning of the numbers
reported on the upper right is the same as in Fig. 5. The thick curve
represents the median value of log (\Ha/\Hb) and the thin curves
represent the 10 and 90 percentiles, meaning that 10\% of the data
points are respectively above or below these curves (the data are binned
according to the abscissa in bins with equal numbers of points).
Such a representation, associated with the relevant statistics, is
believed to give a fair idea of the behaviour of our sample in the
different planes.  Panel $a$ concerns
the galaxy spectral type \texttt{eClass}, panel $b$  the colour
$(u-i)$ as obtained from the SDSS data base.
Panel $c$   the discontinuity at
4000~\AA, $D(4000)$,
defined by the ratio of the
flux from 4050 to 4250~\AA~ to the flux from 3750 to 3950~\AA.
Thus, panels $a$, $b$, and $c$ relate  \Ha/\Hb\ to quantities
  linked to the stellar continuum. Panels $d$, $e$, and $f$
involve respectively the
absolute magnitude $M(r)$,  the surface brightness $S\!B$
defined above and the concentration index $C\!I$
defined as the ratio of \texttt{petroR50}
to \texttt{petroR90} retrieved from the SDSS data base.
Panel $g$ concerns the total radius $R$ of the galaxy in kpc,
computed from the \texttt{petroR90} radius.
Thus, one can consider that panels $d$, $e$, $f$, and $g$ are linked to the
masses and mass distributions of the galaxies.
Panel $h$ concerns the inclination parameter $b/a$ derived
from the SDSS data base.  The fact that the SDSS contains such a large
number of galaxies allows us to infer conclusions from the variation
of observed
properties with galaxy inclination, which was not the case in our previous
studies (SS99, SS01).
Panel $i$ concerns $L({\rm H}\beta)_{\rm corr}$,  the total luminosity
of the
galaxy in \Hb\ (in solar luminosities) after
correction for extinction using the value of $C({\rm H}\beta)$.
Panel $j$ concerns the equivalent width of \Ha. Panels $i$ and $j$
are thus related to the importance of \hii\ regions in the global
galaxy spectra. Finally, panels $k$ and $l$ concern the metallicity
of the galaxies. Among the various metallicity indicators, we have
chosen two which are complementary. The first one,
\Nii/\Ha\ (panel $k$) has been proposed rather recently as a metallicity
indicator in giant \hii\ regions (van Zee et al. \cite{vZ98}, Denicol\'o et
al. \cite{DTT02}).  Its advantages are that it increases monotonically with
metallicity and is independent of reddening, since the two lines it
involves have similar wavelengths. The other one, \Oiii/\Nii\ (panel
$l$) was first proposed by Alloin et al. (\cite{ACJ79}). This indicator
also
varies monotonically with metallicity (it is smaller at large
metallicities) which, in a study of the present kind,  makes it safer to
use  than the traditional (\Oii+\Oiii)/\Hb\ indicator first proposed
by Pagel et al. (\cite{P79}) and still widely in use.
The \Oiii/\Nii\  indicator has one advantage over the \Nii/\Ha\ one:
its larger dynamical range in galaxies. Unfortunately this indicator is
affected by reddening. With the SDSS data, we are able to correct line
intensities for
reddening using the \Ha/\Hb\ ratio, and we use the reddening-corrected
values in our study, assuming that the intrinsic value of \Ha/\Hb\  is
2.9. When using the \Oiii/\Nii\ or \Nii/\Ha\ line ratios, we rejected
the objects where the measured equivalent width in any of these lines
is smaller than 1\AA, in order to avoid spurious values of the ratios
linked to large uncertainties in the measurements of line fluxes.

\begin{figure*}
\centerline{\includegraphics[width=10.9cm]{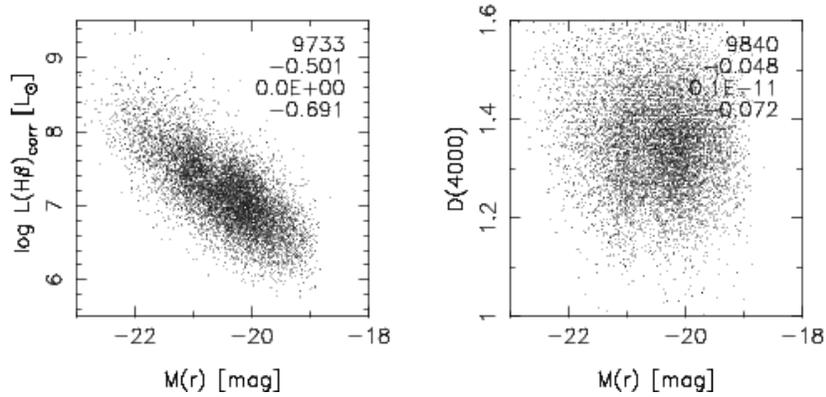}}
\caption{$L({\rm H}\beta)_{\rm corr}$ versus $M(r)$ (left) and $D(4000)$
vesus
$M(r)$. Same layout as Fig. 5.
}
\label{fig7}
\end{figure*}

\subsection{Systematics of  Balmer extinction in galaxies}

Among the investigated relations  between  the
  Balmer extinction  and various parameters of the galaxies, the most
outstanding correlation is the one between \Ha/\Hb\ and the galaxy
spectral types
\texttt{eClass},
for which $r_{S}$ = -0.665, followed by the one with galaxy colour
$(u-i)$ for which $r_{S}$ = 0.630. We find that early-type and redder
galaxies suffer higher Balmer extinction. Of course, without any
modelling, one could be tempted to say that early-type galaxies are
redder not intrinsically, but because they suffer larger reddening.
Such an extreme explanation would of course be unreasonable, since
there are  many arguments showing that early-type galaxies are indeed
composed of older stellar populations (e.g. Sommer-Larsen \cite{S96}, Fioc
\& Rocca-Volmerange \cite{FR97}).
There are, however, indicators of the
characteristics of the stellar population that do not depend on
reddening, like for example the
discontinuity at 4000 \AA, $D(4000)$. This discontinuity is larger
for older and for more metal-rich stellar populations. This is due to
higher
Balmer opacity for lower stellar effective temperatures
dominating the spectra in the first hypothesis, and higher
metal opacity in the second case. It has been argued on
empirical grounds
(Dressler \& Shectman \cite{DS87}) that $D(4000)$ is rather sensitive to
star
formation, but others (e.g. Poggianti \& Barbaro \cite{PB97}) point out
that
metallicity is important. In any case, the correlation we observe
with $D(4000)$ ($r_{S}$ = 0.501) indicates that the Balmer extinction is
higher
for galaxies with older and/or more metal-rich stellar populations.
That the correlation is slightly worse than in panels $a$ and $b$ probably
reflects the fact that $D(4000)$ is not sensitive to reddening.

  We also see a strong correlation  with
$M(r)$ ($r_{S}$ = -0.451) expressing
that the Balmer extinction is higher for more luminous (thus likely more
massive) galaxies. Correlation with  other parameters linked with
galaxy stellar mass and
concentration are present but are less important ($r_{S}$ = -0.298 for
$S\!B$,
$r_{S}$ = -0.182 for $C\!I$ and $r_{S}$ = 0.182 for $R$).

Surprisingly, there is no correlation with $b/a$ ($r_{S}$ = -0.091),
while one
would expect the extinction to be larger for more inclined galaxies. One
interpretation might be that most of the  Balmer extinction
comes from dusty blobs which
are more or less uniformly distributed in the galaxy, so that the
extinction of
the integrated flux does not depend on inclination. However, such an
interpretation is not compatible with the fact the near-infrared
colours of spiral galaxies depend on inclination (Masters et al.
2003). We will come
back to this point later.

There is a weak anticorrelation with   \EWHa\ ($r_{S}$ = -0.194).
\EWHa\ is a star formation history indicator (Kennicutt 1998) and the
observed anticorrelation suggests that the Balmer extinction is
higher in galaxies with a
larger proportion of old stellar populations. On the other hand, the
correlation
with the total \Hb\ luminosity
($r_{S}$ = 0.607) suggests that the Balmer extinction is larger for
galaxies with a larger
total amount of young stars.

Finally, correlations with the metallicity
indicators are fairly important (panels $k$ and $l$):  $r_{S}$ = 0.418 for
\Nii/\Ha\
and $r_{S}$ = -0.449 for  \Oiii/\Nii, suggesting that metallicity and
Balmer
extinction are strongly correlated.

\subsection{Is the metallicity of galaxies the only driver of
Balmer extinction? }
It is known that many properties of galaxies are correlated in the
Hubble sequence of galaxies (Roberts \& Haynes \cite{RH94}).
Early-type galaxies tend to be more massive, composed of
older stellar populations and more metal-rich
(see e.g. Zaritsky et al. \cite{ZKR94}, Kennicutt \cite{K98},
Fioc \& Rocca-Volmerange \cite{FR97}).
Could it be that the main parameter responsible
for the trends mentioned above between $C({\rm H}\beta)$ and other
galaxy parameters is simply metallicity?

Let us assume for simplicity that $C({\rm H}\beta)$ can be written as:
\begin{equation}
C({\rm H}\beta) = f(Z) g(X),
\end{equation}
where $f(Z)$ is a function of metallicity and $g(X)$ is a function of
other parameters.

We know that \Nii/\Ha\
and \Oiii/\Nii\ are metallicity indicators and that they vary
monotonically with metallicity. The problem is that we do not know
the exact relation between the value of these ratios and the average
metallicity of  the interstellar medium in the galaxy. It is
not possible to directly use the
calibrations proposed for giant \hii\ regions (which in themselves
are already a matter of debate, see Pilyugin \cite{P03}) because, as
mentioned by SS01,
spectra of galaxies are affected by metallicity gradients
and by the contribution of diffuse ionized regions to the spectrum.
This can be illustrated as follows. In the sample of  spiral galaxies
studied by Zaritsky et al. (\cite{ZKR94}) the average metallicities
vary from
12 + log (O/H) = 8.3 to 9.3 when the galaxy absolute B
magnitudes $M_{B}$ range from -17 to -21. The slope of the relation between
log O/H and $M_{B}$ in their Fig. 10 is roughly -0.25 (but there is
substantial dispersion).
If we would apply the calibration of Van Zee et al. (\cite{vZ98}) (
12 + log\,(O/H) = 1.02 log\,(\Nii/\Ha) + 9.35), the
corresponding \Nii/\Ha\ values would range between -1.04 and -0.06
and the slope between log \Nii/\Ha\ and $M_{B}$ would again roughly be
-0.25. Figure 7 shows the relation between
log\,(\Nii/\Ha)  and  $M_{B}$  in our sample (the value of $M_{B}$
has been estimated using the transformations between the SDSS
photometric system and the standard $UBV$ system following Smith et
al. \cite{STK02}). Most of the observed values of
log\,(\Nii/\Ha) are between $-0.7$ and $-0.3$ even for the galaxies with
the lowest luminosities in our sample. This is likely due to the fact
  that in spectra of galaxies
both the radial metallicity gradients and the contribution of
emission from the diffuse ionized
medium would tend to enhance the \Nii/\Ha\ ratio.

What we can do, however, is study the function
$C({\rm H}\beta)$/(\Nii/\Ha)$^{a}$
where $a$ is determined empirically so as to remove any dependence of
this function  on \Nii/\Ha,
  implying that $C({\rm H}\beta)$/(\Nii/\Ha)$^{a}$ is independent of
metallicity.
  We find that $a$ is approximately equal to 0.9 in our sample. While
this procedure is very
schematic and relies on
assumptions, the fact that  we find $C({\rm
H}\beta)$/(\Nii/\Ha)$^{0.9}$ to be  independent of the other
metallicity indicator, \Oiii/\Nii, gives us confidence in our
approach. (Note that the value of $a$ depends on the definition of
our sample and should not be used for other samples without caution).
Figure 8 shows the same plots as Fig. 6, with log (\Ha/\Hb)
   now replaced by $C({\rm
H}\beta)$/(\Nii/\Ha)$^{0.9}$.
  We see that $C({\rm
H}\beta)$/(\Nii/\Ha)$^{0.9}$ (which can be assimilated to $g(X)$ in Eq.
2) is strongly correlated
with  the galaxy spectral type \texttt{eClass} ($r_{S}$ = -0.436) and
  colour $(u-i)$ ($r_{S}$ = 0.394) and also with $D(4000)$ ($r_{S}$=0.340)
although less tightly than $C({\rm H}\beta)$. The strongest correlation
in this
figure is with $L({\rm H}\beta)_{\rm corr}$ ($r_{S}$ = 0.512).
However, we show in Fig. 9, that $L({\rm H}\beta)_{\rm corr}$ is very
strongly
correlated with $M(r)$ ($r_{S}$ = -0.691) so that we conclude that an
important  driver of the  Balmer extinction is actually the
total galaxy luminosity (or mass).
On the other hand, $D(4000)$ is not correlated with $M(r)$ as seen
in Fig. 9, so the observed correlation between $C({\rm
H}\beta)$/(\Nii/\Ha)$^{0.9}$  and $D(4000)$ implies that the presence of
old
stars is in itself also a driver of the  Balmer extinction.

We also note in Fig. 8 (panel $h$) that $C({\rm
H}\beta)$/(\Nii/\Ha)$^{0.9}$ is
linked with galaxy inclination ($r_{S}$ = -0.200). One explanation
that can account
for this as well as for the absence of correlation between  \Ha/\Hb\ and
$b/a$ is that \hii\ regions are rather close to the galactic plane of
galaxies (which is in better agreement with the observed height scale of
\hii\
regions in nearby galaxies), and that the increased path length of the
attenuating dust in inclined galaxies is roughly compensated by abundance
gradients: when a galaxy is inclined, the innermost regions, that are more
metal-rich and therefore most likely dust-rich, are more obscured than the
outermost \hii\ regions, which suffer less extinction at optical
wavelengths. The
compensating effect of the metallicity gradients at near-infrared
wavelengths is
far less important due to smaller optical depths at these
wavelengths, which explains why Masters et al. (2003) do find a
correlation between near-infrared colours and inclination.  We have
tested by
simple toy models that such an explanation can be viable.

\section{Discussion}

We have thus found (confirming the results of SS99 and SS01) that the
Balmer extinction of galaxies decreases steadily from early- to
late-type
spirals (whether classified according to their spectral type or
according to their colour). We have also shown that there is a
direct dependence of Balmer extinction on metallicity, and also
on the age of the stellar populations and on
the total galaxy luminosity.
This last statement is consistent
with the finding of Wang \& Heckman (\cite{WH96}),
based on relations using far-ultraviolet
and far-infrared fluxes, that ``the optical depth of normal galactic
disks increases with
galaxy luminosity''. On the other hand, the dependence we find
on galaxy type seems to
contradict the general opinion summarized by Calzetti (\cite{C01})
and the recent result by  Kauffmann et al.  (2003) who conclude that
``galaxies with the youngest stellar populations are the most attenuated
by dust''.
   While previous studies were sometimes based on small samples, the
result of Kauffmann et al. (2003) comes from a model fitting of the
continua of  $10^{5}$ galaxies from the SDSS. It should be noted that what
they
measure with this elaborate procedure is the attenuation of the 
\emph{stellar} light.  Our work concerns
the extinction of \emph{nebular} light and
the determination of the Balmer extinction is straightforward.  The
interpretation in terms of global opacity of galaxies is however not
simple:
with  \Ha/\Hb\  we measure  some sort of average
opacity of the zones dimming the light from \hii\ regions (excluding 
the most opaque ones). Still, it is
striking to see how well $C({\rm H}\beta)$ correlates with the galaxy
spectral type. In this section, we present a possible
phenomenological interpretation of our results, and show that our
findings can actually be reconciled with the results of Kauffmann et 
al. (2003).

\begin{figure}
\centerline{\includegraphics[width=5.cm]{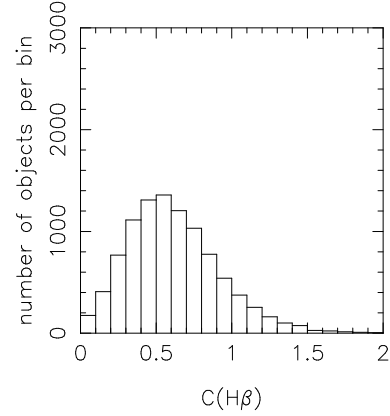}}
\caption{Histogram of the values of the Balmer extinction at \Hb,
$C(\Hb)$ in our sample of galaxies.
}
\label{fig8}
\end{figure}

\subsection{A phenomenological interpretation of the observed Balmer
extinction trends in our sample}
Let us write $C({\rm H}\beta)$
as the following product:
\begin{equation}
C({\rm H}\beta) = n_{\rm d} l \sigma_{\rm d} =
\frac{n_{\rm d}}{n_{\rm M}} \frac{n_{\rm M}}{n_{\rm H}}
\frac{n_{\rm H}}{n_{*}} n_{*} l \sigma_{\rm d}
\end{equation}
where $n_{\rm d}$ is average number of dust particles per unit volume in
a galaxy,  $n_{\rm M}$ is average  number of metallic atoms (in any form)
per unit volume,
   $n_{\rm H}$  is the average number of hydrogen particles per unit volume,
$n_{*}$ is the average number of stars per unit volume, $l$ is the
geometrical thickness of the region responsible for the optical
extinction,  and $\sigma_{\rm d}$ is the typical extinction
cross section of dust grains.

The common view that the opacity should increase from early- to
late-types is comforted by such arguments that  ``galaxies with young
stars contain more gas and hence more dust than galaxies with old
stellar populations'' (Kauffmann et al. \cite{K03}).
On the other hand, early-type galaxies are more metal-rich and on
average more massive than late-type galaxies, so it is not necessarily
surprising that the extinction is actually larger for late-type galaxies.
In Sect. 3.2 we have given arguments to say that the Balmer
extinction is also
determined by the mean age of the stellar population. This could
indicate that in late-type galaxies $n_{\rm d}/n_{\rm M}$ is larger than in
early-type ones. Such a view seems indeed to be supported by recent
models for
the evolution of dust in galaxies (Hirashita \cite{H99}) which take into
account the processes of formation and destruction of dust (note however
that there are presently many assumptions and uncertainties in such
models, see e.g. Dwek \cite{D98} or Edmunds \cite{E01}). In these models,
condensation in cool stellar winds from low-mass stars is an
important source of dust production.

We note that 75\% of the galaxies of our sample have a
measured $C({\rm H}\beta)$
between 0.3 and 0.9, as seen from the histogram shown in Fig. 10.
On the other hand, judging from the metallicities derived for galaxies of
luminosities similar to those of our sample (Zaritsky et al. \cite{ZKR94},
Charlot et al. \cite{CKL02}), the  metallicity
range in our sample is likely higher than just a
factor of three. This
suggests that factors other than metallicity act to reduce the observed
range in $C({\rm H}\beta)$. Obviously, $n_{\rm H}/n_{*}$ is a good
candidate,
since it decreases from late- to early-type galaxies (Roberts \&
Haynes \cite{RH94}).

On the other hand $ n_{*} l$, which can be assimilated to the stellar
surface
density, decreases from early- to late-types.

The last factor in Eq. (3) is  $\sigma_{\rm d}$, and one might
expect some systematic effects if the grain size distribution depends on
the processes for grain growth or destruction that could have
different relative importances in galaxies of
different types.

In conclusion, dust extinction in galaxies involves many
factors, and our finding that $C({\rm H}\beta)$ increases from late-
to early-types can easily be accounted for, at least qualitatively,
within our present-day understanding of galaxies and their
constituents.

\begin{figure}
\centerline{\includegraphics[width=5.13cm]{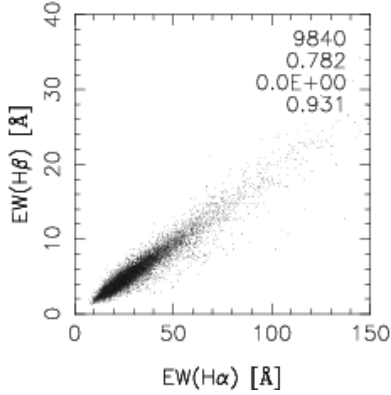}}
\caption{The relation between \EWHb and \EWHa\ in our sample of
galaxies. The layout of the figure is the same as for Fig. 5.
}
\label{fig9}
\end{figure}

\subsection{The extinction of stellar light versus the extinction of
nebular
light}

As emphasized above, what we measure with $C({\rm H}\beta)$ is
actually the reddening of the nebular light (called $C_{l}$ in SS01).
On the other hand, what is determined by
Kauffmann et al. (2003) is  the reddening of the stellar light
(called $C_{c}$ in SS01). We now examine the relation between
these two quantities in our sample of galaxies.

It had already been noted by SS99 and SS01, on much smaller samples of
galaxies, that  \EWHa\  and \EWHb\ correlate extremely well (see also
Kennicutt 1992). This
implies that the difference between
the extinction of the stellar continuum and that of the nebular
emission is strongly linked
to the colours of the galaxies, being larger for redder
galaxies.  (see Eq. 5 in SS01).
In Fig. 11 we show the values of  EW(H$\beta$) as a function of
EW(H$\alpha$) in our sample
of galaxies. The correlation is extremely strong, with $\tau_{K} =
0.782$ and $r_{S} = 0.931$. Assuming that the relation between these
emission line equivalent widths is EW(H$\beta$) = $A \times$ EW(H$\alpha$)
(the same model as adopted in SS01), an ordinary least square bisector
linear
fitting (Isobe et al. 1990) gives $A = 0.185 \pm 0.001$.
Considering only objects for which EW(H$\alpha$) $>$ 20 \AA\ the result
is the
same, $A = 0.185 \pm 0.001$. These results may be compared with
$A = 0.194 \pm 0.011$ and $A = 0.245 \pm 0.007$, presented in SS99 and
SS01,
respectively. Our value   $A$ is consistent with that obtained by SS99
but
not with that obtained by SS01. The cause of this discrepancy is not
clear;
it might be due to differences in the spectral resolution, because
the spectral resolution of the SDSS spectra is almost half that of the
NFGS spectra (Jansen et al. \cite{JFFC00a}, \cite{JFFC00b}) used by SS01.

 From Eq. (5) in SS01 and the value of $A$, we derive
that
\begin{equation}
C_{c} = C_{l} - 0.81 - 2.99 ~ {\rm log}\, \frac {F_{c}^{o}
({\rm H}\alpha)}{F_{c}^{o}({\rm H}\beta)},
\end{equation}
where $F_{c}^{o}$(H$\alpha$) and $F_{c}^{o}$(H$\beta$) are the
intrinsic (i.e. not affected by extinction) stellar
fluxes in the continuum adjacent to \Ha\ and
\Hb\ respectively. We can use the dust-free spectrophotometric models
of Barbaro \& Poggianti (1997) to relate the value of
$F_{c}^{o}$(H$\alpha$)/$F_{c}^{o}$(H$\beta$) to the values of
$D(4000)$ and \EWHa. Using panel $j$ of our Fig. 6 we can relate
\EWHa\ to log (\Ha/\Hb) (by taking the median value of \Ha/\Hb\ for a
given \EWHa) and therefore to  $C_{l}$. With the help of Eq. (4)
we thus find that $C_{c}$
goes from  about  0.02 for $D(4000) = 1.7$ to
  about  0.2 for $D(4000) = 1.3$.
  This goes in the same direction as the results of  Kauffmann et al. (2003)
who find that the median value of $A_{z}$, which is approximately
equal to our $C_{c}$, roughly goes
from 0.2 at $D(4000) =
1.7$ to 0.6 at $D(4000) =
1.3$. That the numbers are not exactly the same as the ones found by our
analysis is not necessarily a worry given the dispersion in the
observational
points (both here and in the work of Kauffman et al. 2003) and given
that the
definition of the derived extinction is not exactly the same.

As for $C_{l}$, it can be evaluated using Fig. 6c and Eq. 1. We find
$C_{l}$   $\simeq$
0.8 for $D(4000) = 1.7$ and $C_{l}$ $\simeq$ 0.3 for $D(4000) = 1.3$
  We note that $C_{c}$ is smaller than  $C_{l}$ at both extremes of
the spectral type range, which can be interpreted as due to the fact
that dust is more concentrated (and thus more opaque to radiation) in
molecular clouds associated with \hii\ regions than in the diffuse
interstellar medium. One may wonder why $C_{c}$ increases from early-
to late-type spirals while
$C_{l}$ decreases. The answer to this may be related to the fact
that the stellar light from early-type galaxies is dominated by the
bulge and to a specific distribution of dust resulting from dynamical effects.
Advanced 3D-modelling of the star, dust and gas
distribution in galaxies would be needed to test  any interpretation
of our empirical result.

\begin{figure*}
\centerline{\includegraphics[width=10.9cm]{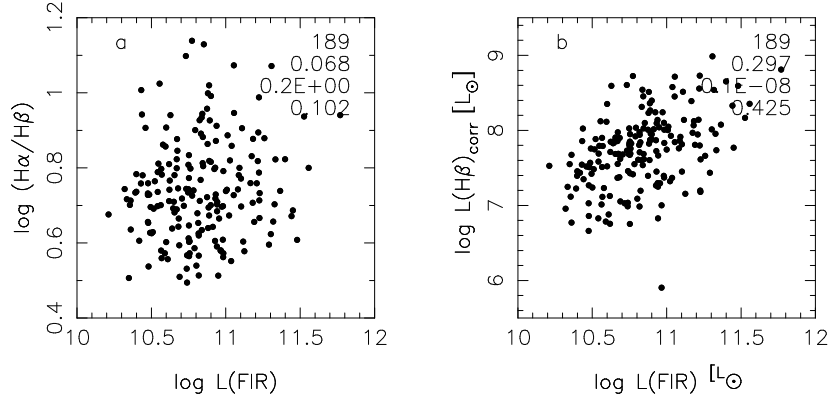}}
\caption{Galaxies from our sample detected by IRAS (see Sect. 5).
Plots of log (\Ha/\Hb) vs. log $L(\rm FIR)$
(panel $a$) and log $L(\Hb)$ (corrected for Balmer extinction)  vs 
log $L(\rm FIR)$ (panel b).
}
\label{fig10}
\end{figure*}

\begin{figure*}[t]
\centerline{\includegraphics[width=16cm]{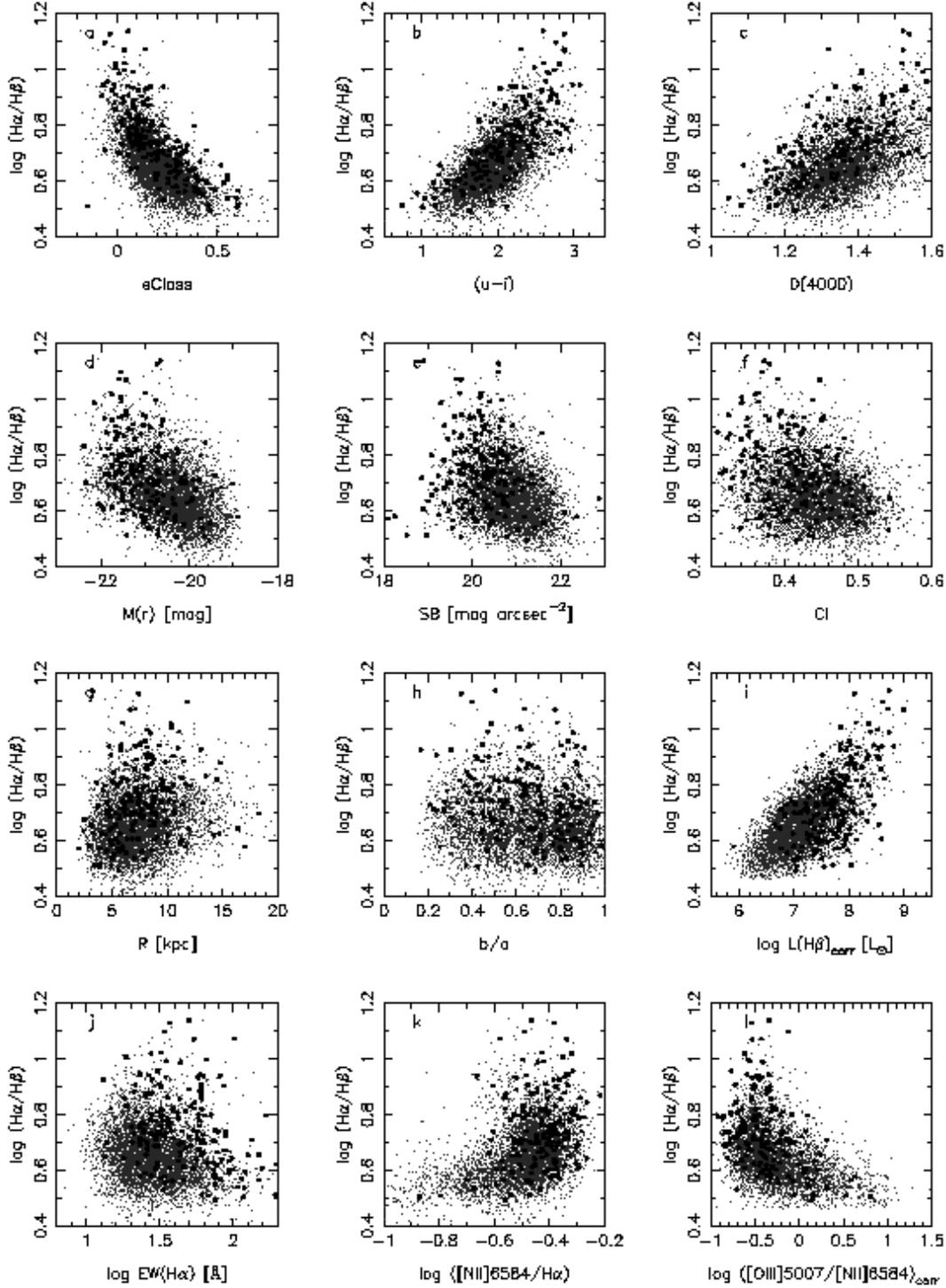}}
\caption{Same diagrams as in Fig. 6, with the galaxies detected by
IRAS represented with large circles.
}
\label{fig11}
\end{figure*}

\section{The relation between  $C({\rm H}\beta)$ and infrared
luminosity in our sample}

Another indicator of the presence of dust in galaxies is their far-infrared
radiation attributed to the emission from dust grains heated by
stellar radiation (e.g. Inoue \cite{I02}). Kewley et al.\,(\,2002),
using 81 galaxies with
relevant data from the NFGS sample, noted the existence of a very strong
correlation between the far-infrared luminosity $ L({\rm FIR})$ and the
\Ha\
luminosity
(corrected for reddening using the observed \Ha/\Hb\ ratio). They
find a Spearman rank correlation coefficient of $r_{S}$ = 0.98. They do
not plot the relation between \Ha/\Hb\ and $ L({\rm FIR})$, but this
plot is
shown by Wang \& Heckman (\cite{WH96}) for about 30 galaxies from the
Kennicutt (\cite{K92}) atlas with relevant data, and they find a
correlation
to within better than  a 0.1\% level of confidence (the
correlation coefficient from a
standard linear regression is 0.63).

It was tempting to produce such diagrams for our sample of galaxies
using the IRAS data base.
We have searched the IRAS Faint Source Catalogue (Moshir et al. \cite{M89})
and the IRAS Point Source Catalogue (IPAC \cite{IPAC86})
for far-infrared counterparts of galaxies from our sample. Since the IRAS
beam size at 60\,$\mu$m is 1$\farcm$5 and the IRAS uncertainty in
position is
about 30$\arcsec$, we used a conservative circular window of 1$\arcmin$
radius
around the SDSS positions of our sample galaxies to search for far-infrared
sources. Note that we also constituted a subsample of IRAS
galaxies using a search window of  30$\arcsec$ instead of
1$\arcmin$; this subsample is twice smaller but the conclusions drawn
below are
not altered.
Since an estimate of the total infrared luminosity requires
IRAS fluxes  at both 60\,$\mu$m and 100\,$\mu$m ($f_{60}$ and
$f_{100}$,
respectively), we kept only those galaxies with available  fluxes in both
bands and IRAS quality flags greater than 1. About 80 galaxies were
common to both the IRAS Point Source Catalogue and the IRAS Faint
Source Catalogue. For those objects, we retained the flux values
given by the Faint Source Catalogue, which have greater quality.

This search resulted in 189 galaxies.
We have computed $L({\rm FIR})$ using the formula from Thuan \&
Sauvage (\cite{TS92}):
\begin{equation}
  L({\rm FIR}) = 3.95\times10^{5}(2.58 f_{60} + f_{100} )
D_{\rm L}^{2}
\end{equation}
where  $L({\rm FIR})$ is in solar units,  $f_{60}$ and $f_{100}$ are in
Jy and
$D_{\rm L}$ is the cosmological luminosity distance of the galaxy in Mpc.
In most cases, $ L({\rm FIR})$
$ > 2 \times 10^{10}$ L$_{\odot}$.  Note that this lower limit actually
corresponds to the high luminosity tail of the luminosities of spiral
galaxies
detected by IRAS (Thuan \& Sauvage 1992). This stems from the fact that our
samples contains only galaxies at redshifts larger than 0.05, and at such
distances only luminous infrared galaxies can be detected.

Figure 12 plots the values of log\,(\Ha/\Hb) (panel $a$)  and log\,$L({\rm
H}\beta)_{\rm corr}$ (panel $b$) versus log\,$ L({\rm FIR})$.
We do not see any correlation between  log\,(\Ha/\Hb) and $ L({\rm
FIR})$ (the
probability associated to the Kendal statistics is as large as 0.2).
A correlation between $L({\rm
H}\beta)_{\rm corr}$ and  $ L({\rm FIR})$ is present in our sample, but
much
weaker than found by Kewley et al. (2002). It
thus turns out that
   correlations between optical and infrared properties become much
   less conspicuous or even disappear when restricting the sample to 
luminous infrared
   galaxies\footnote{We use the term ``luminous infrared galaxies'' for
   convenience, although it is
   traditionally reserved for galaxies with  $ L({\rm FIR}) > 10^{11}$
   L$_{\odot}$ (Sanders \& Mirabel \cite{SM96}).}.
   Note that in the samples of Wang \& Heckman (\cite{WH96}) and
   Kewley et al. (\cite{KGJD02}) galaxies with
$ L({\rm FIR}) > 2 \times 10^{10}$ L$_{\odot}$
   are only a handful and do not show a clear correlation either.

It is also interesting to see how our luminous infrared galaxies
compare with our entire sample of galaxies. For this, we show
in Fig. 13 the same diagrams as in Fig. 6, with the luminous infrared
galaxies now indicated by large circles.
  We see that in all the diagrams the galaxies detected by IRAS are found in
the entire domain covered by the SDSS galaxies except the regions
corresponding
to the least luminous galaxies (which cannot be detected by IRAS at
the redshits
of our SDSS sample). The distributions of the luminous infra-red
galaxies is
very similar to that of our entire set of SDSS galaxies, except that is is
slightly skewed towards the highest values of \Ha/\Hb.

The comparison of our SDSS sample with IRAS data  raises two
questions. Why were
only 2\% of our sample of SDSS galaxies detected by IRAS?
Why are those IRAS detected galaxies found at any value of \Ha/\Hb?
Concerning the first question, we have already noted that, because of
the limit
imposed on the redshift in our SDSS sample, only luminous infrared
galaxies can
be identified as IRAS sources. An IRAS source is characterized by the
presence
of warm dust (50-200\,K), which lies relatively close to the stars. Such a
situation is encountered in zones where star formation has occured very
recently, and which have not yet been swept out by stellar winds. In a
galaxy,
the \hii\ regions that emit the observed H Balmer lines have ages
between 1 and
10\,Myr, but the \hii\ regions that will emit most efficiently in the
IRAS bands
are younger than that. This explains, at least qualitatively, why only a
small
proportion of the galaxies from our SDSS sample have been detected by IRAS.
These are the ones that happen to have a larger number of zones of very
recent
star formation. Such galaxies are more likely found among more massive
galaxies
at a given spectral type, but may be found at any galaxy spectral type.
Since
the IRAS emission probes these extremely young star forming regions, 
there is no
reason why it should show any correlation with the global optical 
properties of the
galaxies, which answers the second question.

\section{Summary, implications and prospects}

We have used the observations from the First Data
Release of the SDSS to examine a sample of normal galaxies (as
opposed to galaxies with an active nucleus) and to investigate  how
the Balmer extinction $C({\rm H}\beta)$ (i.e.
  the extinction at the wavelength of \Hb\ derived from the \Ha/\Hb\
emission line ratios)
relates with
other global properties of the galaxies.
Our selection criteria to build up the sample resulted in a data set
of 9840 galaxies with adequate data. All these galaxies are at
redhifts larger than 0.05 to avoid strong aperture
effects.

Our main findings are the following:
\begin{enumerate}
\item $C({\rm H}\beta)$  is linked with the galaxy
spectral type and colour, decreasing from early- to late-type spirals.
\item $C({\rm H}\beta)$ increases with increasing metallicity
\item $C({\rm H}\beta)$ is, probably, also affected by the
age of the stellar population, being larger in the case of older
stellar populations.
\item $C({\rm H}\beta)$ depends on galaxy masses.
\item  The extinction of the stellar light is correlated with
both the extinction of the nebular light and the intrinsic galaxy
colours, resulting in a trend with galaxy colour that may be opposite
to the trend of $C({\rm H}\beta)$.
\end{enumerate}

The present work thus confirms the conclusions of our previous studies
(Sodr\'e \& Stasi\'{n}ska \cite{SS99} and Stasi\'{n}ska \&
Sodr\'e \cite{SS01}), which
were based on much smaller samples and used data with lower spectral
resolution (Kewley et al. \cite{KGJD02}, using the same sample as
SS01, also found that early-types in that
sample are more heavily reddened than late-types).

  Compared to our previous studies, the large number of galaxies in the SDSS
sample allows us to investigate issues related to the inclination of
galaxies.

The fact that  $C({\rm H}\beta)$
correlates so well with other properties of galaxies  is
remarkable, given that the extinction, especially in
late-types, is known to be not uniform across the face of
galaxies (e.g. Beckman et al. \cite{BPK96}).

We have cross-correlated our sample of SDSS galaxies with
the IRAS data base in order to investigate any relationship
between $C({\rm H}\beta)$ and total infrared luminosity of the
galaxies. Due
to the lower redshift limit imposed to our sample and to the detection
limit of IRAS, such a procedure selected only luminous infrared
galaxies. We found that correlations that were shown by other authors
to exist between optical and infrared properties of galaxies
disappear when restricting to luminous infrared galaxies. We also found
that the optical properties of the luminous infrared galaxies in
our SDSS sample are very similar to those of our entire sample of
SDSS galaxies.

  We have proposed a phenomenological interpretation of our
findings.  We suggest
that the main driver of the  Balmer extinction of galaxies is their
mass, combined with their metallicity and presence of old stellar
  populations.
The infrared luminosity of the galaxies as determined by IRAS, which is
attributed to radiation from hot stars reprocessed by dust grains,
  samples the regions with {\it the most recent episodes} of star 
formation, and is
not
connected with the Balmer extinction.
Obviously, detailed modelling of the spectral
light from galaxies taking into account the effects of dust and using
a complete
code such as GRASIL (Silva et al. 1998, see also
http://web.pd.astro.it/granato/grasil/grasil.html)
  is needed for a
deeper understanding of the empirical relations we have found. This is
not an easy task, however, since as noted by Witt at al. (1992) and
Witt \& Gordon (2000),
equal amounts of dust in different configurations may produce very
different reddening and attenuation effects.
In any case, future models of the integrated light from galaxies
including the effects of dust should also aim at reproducing the
correlations
we have shown.

An important outcome of our study is to open the way for an
improved correction for extinction in the determination of such
parameters as the global star formation rate in galaxies or their
total stellar masses. For normal galaxies, the global star formation
rate can be obtained from the total \Ha\ luminosity corrected for
extinction using the extinction derived from the \Ha/\Hb\ emission
line ratio (keeping in mind the reservations expressed e.g. by
Hirashita et al. 2003). If observations do not allow one to determine the
\emph{Balmer extinction}, one can make use of e.g. the observed
galaxy colour or the $D(4000)$ parameter to obtain an estimate of
statistical value since we have shown that all these quantities are
correlated. On the contrary, the total stellar mass can be estimated
from the observed stellar fluxes of the galaxy after correcting for
\emph{stellar extinction}.  This should be done with a proper model
fitting of the observed continuum as in Kauffmann et al. (2003).
However, the strong correlation
that we have found empirically between  stellar extinction, Balmer
extinction and  galaxy colours can provide a basis for a
statistical method to determine the total masses of galaxies.
These aspects will be developed in future work and should be
important especially for the study of galaxies at intermediate and
high redshifts.

\begin{acknowledgements}

This work has benefited from grants from the PICS
franco-br\'{e}silien, from the jumelage France-Pologne and
from the Observatoire de Paris. A.M. and L.S. acknowledge support from
FAPESP and CNPq; L.S is also grateful to CCINT/USP.
The Instituto de Astronomia, Geof\'{\i}sica e
Ci\^encias Atmosf\'ericas da USP has provided hospitality
to G. S. and the Observatoire de
Paris to L. S. and R. S.  All the authors wish to thank the team of the
   Sloan Digital Sky Survey (SDSS) for their dedication to a project which
   has made the present work possible.
The Sloan Digital Sky Survey is a joint project of The University of
Chicago, Fermilab, the Institute for Advanced Study, the Japan
Participation
Group, the Johns Hopkins University, the Los Alamos National Laboratory,
the
Max-Planck-Institute for Astronomy (MPIA), the Max-Planck-Institute for
Astrophysics (MPA), New Mexico State University, Princeton University, the
United States Naval Observatory, and the University of Washington.
Funding for the project has been provided by the Alfred P. Sloan
Foundation,
the Participating Institutions, the National Aeronautics and Space
Administration, the National Science Foundation, the U.S. Department of
Energy,
the Japanese Monbukagakusho, and the Max Planck Society. Thanks are
also due to the referee whose comments led to a substantial
clarification of the paper.

\end{acknowledgements}

\end{document}